\documentclass[doublecol]{arXiv}

\title{On the mechanisms of precipitation of graphene on nickel thin films}
\shorttitle{Mechanisms of precipitation of graphene on Ni} 

\author{L. Baraton\inst{1}\footnote{Now at Laboratoire de G\'{e}nie \'{e}lectrique de Paris, CNRS UMR8507; SUPELEC; UPMC Univ. Paris 6; Univ. Paris-Sud - 11 rue Joliot Curie, Gif-sur-Yvette 91192, France}  \and Z. B. He\inst{1}\footnote{Now at EMAT, University of Antwerp - Groenenborgerlaan 171, Antwerp B-2020, Belgium}  \and C. S. Lee\inst{1} \and C. S. Cojocaru\inst{1} \and M. Ch\^{a}telet\inst{1} \and J.-L. Maurice\inst{1}\footnote{corresponding author: jean-luc.maurice@polytechnique.edu} \and Y. H. Lee\inst{2} \and D. Pribat\inst{2}\footnote{corresponding author: didier53@skku.edu}}
\shortauthor{L. Baraton \etal}

\institute{                    
  \inst{1} LPICM CNRS \'{E}cole polytechnique - Palaiseau 91120, France
  
  \inst{2} Department of Energy Science, Sungkyunkwan University - Suwon 440-746, Korea

}

\abstract{
Growth on transition metal substrates is becoming a method of choice to prepare large-area graphene foils. In the case of nickel, where carbon has a significant solubility, such a growth process includes at least two elementary steps: (1) carbon dissolution into the metal, and (2) graphene precipitation at the surface. Here, we dissolve calibrated amounts of carbon in nickel films, using carbon ion implantation, and annealing at 725$^{\circ}$C or 900$^{\circ}$C. We then use transmission electron microscopy to analyse the precipitation process in detail: the latter appears to imply carbon diffusion over large distances and at least two distinct microscopic mechanisms.}

\begin{document}

\maketitle

\section{Introduction}
The catalytic growth of graphene on a transition metal substrate has appeared during the last two years as the method delivering the best compromise between cost and high quality. A remarkable example is that of graphene obtained by chemical vapor deposition (CVD) on copper foils, under the form of 75-cm-diagonal sheets\cite{Bae2010}. The CVD technique, which appears as the most advanced growth method to date, has actually been used to grow graphite/graphene on polycrystalline metal films or foils for decades\cite{Robertson1969, Fedoseev1979}; it delivers graphene by thermally decomposing carbon-based gaseous precursors such as methane or ethylene\cite{Yu2008, Reina2009, DeArco2009, Kim2009, Li2009}. Once synthesized on the metallic substrate, graphene can be separated by selectively etching the supporting metal layer and can then be transferred to an arbitrary substrate. The case of copper is special among the possible substrates, as carbon exhibits a low solubility in that metal, so that its surface likely acts directly as a catalyst for the growth\cite{Li2009}. Nickel, the object of the present study, exhibits on the contrary a significant carbon solubility, and in addition a high carbon diffusivity\cite{Lander1952}. Quite surprisingly, given that carbon invades the bulk of nickel films during a standard growth, this catalyst allows one to grow graphene layers an order of magnitude faster than copper\cite{Reina2009, Kim2009}. CVD or other methods used to bring carbon atoms on or into the nickel (solid carbon-based films, ion implantation\cite{Garaj2010, Baraton2011}) essentially serve as a means of carbon doping of the nickel films, so that all processes include at least the two steps: (\textit{i}) dissolution of the (released) carbon atoms into the metal at high temperature (700-1000$^{\circ}$C) and (\textit{ii}) crystallization of carbon atoms onto the metal surface to form graphene. The latter step may take place during the high temperature period of the treatment or during cooling, the driving force being a combination of equilibrium surface segregation and precipitation due to saturation of the solid solution\cite{Yu2008, Reina2009, Kim2009, Sutter2008, Amara2008, Coraux2008, Amara2009}. The same phase separation effect takes place when other transition metals are carbon-doped\cite{Sutter2008, Starr2006, Makarenko2007, Bolotov1998}. 

Regarding nickel more precisely, the exponential variation of solubility with temperature associated with a diffusivity that remains fast over a large temperature range (see below), make graphene growth on that metal drastically dependent on the details of thermal history\cite{Kim2009}. It is thus particularly timely to experimentally investigate the graphene segregation/precipitation mechanisms by isolating the different stages of heat treatment. In this work, in an attempt of simplification of the growth process, we have introduced carbon atoms into the metal by using ion implantation (Io-I)\cite{Baraton2011}. This allowed us to eliminate some of the uncertainties concerning how and how much carbon is incorporated into the metal. In other words, we know exactly (as we can pre-determine it) the carbon content into the metal before annealing at high temperature and more importantly, we know that, before annealing, the in-plane carbon concentration is uniform throughout the sample. Io-I is a routine process developed by the semiconductor industry over the past few decades\cite{Ziegler2010} and it is now being explored for growing functional graphene films on nickel\cite{Garaj2010, Baraton2011}. One of the advantages of Io-I is that it allows one to precisely control the amount of impurity introduced in the host material/substrate. 

\section{Carbon precipitation in Ni}
Here, we precipitate carbon at the surface of a nickel film by high temperature treatment of a given implanted dose. Let us start by analysing the solubility of carbon in Ni and the way it precipitates. Although the results of a given thermal treatment will depend both on static and kinetic parameters, it is most important to know the reference thermodynamic (equilibrium) state that corresponds to a given temperature-carbon concentration couple. Let us first recall some general definitions. I Equilibrium segregation is a compositional heterogeneity occurring at lattice discontinuities (\textit{e.g.}, surfaces) under thermal equilibrium conditions and corresponding to a one-phase domain of the phase diagram; equilibrium segregation minimizes the free energy of the unsaturated solution. II Precipitation corresponds to a classical phase separation phenomenon in agreement with the thermodynamic phase diagram. In the case of the 2D graphene phase, one sees that there is an interesting point at the crossroads of precipitation and segregation. That point has been quite thoroughly studied by Blakely and coworkers in the seventies\cite{Shelton1974, Eizenberg1979, EizenbergJCP1979}, and more recently by Gamo \etal\cite{Gamo1997}. The former authors have analyzed nickel surfaces in situ during thermal treatments of carbon-doped nickel samples around the precipitation temperature. They were able to evidence a specific 2D crystallization regime, corresponding to the formation of a single graphene layer, which they called the ''segregation'' regime (although it corresponds to a phase transition contrary to the classical equilibrium segregation defined above). That specific ''segregation'' phenomenon takes place at temperatures above the standard precipitation temperature, \textit{viz.} at carbon concentrations below the classical solubility limit. They found that this phenomenon occurs preferentially on (111), (311) and (110)-oriented Ni surfaces\cite{EizenbergJCP1979}. The formation of carbon films on nickel upon cooling thus follows two stages: (\textit{i}) graphene crystallization on the one hand, which Blakely and co-workers called ''segregation'', taking place at $T_S$, associated with a solubility $S_S$, and (\textit{ii}) graphite precipitation on the other hand taking place at $T_P$, approximately 100 K below $T_S$ for a given carbon concentration, which is the classical “precipitation” corresponding to the phase diagram solubility $S_P$. 

The expression for solubility is $S_P = S_{P0}  \exp(H_P /kT)$ (in atoms cm$^{-3}$), where $S_{P0}$ is an entropic pre-factor related to the density of sites where solute atoms sit (interstitial here), $H_P$ the heat of precipitation and $k$, Boltzmann$^{\prime}$s constant. Lander and co-workers\cite{Lander1952} have experimentally derived these quantities in the 1000-1650 K temperature range; they have found $S_{P0} = 5.33\times10^{22}$ atoms cm$^{-3}$ and $H_P$ = -0.421 eV (their data in weight percent and kelvins have been translated into atomic concentrations and eV per atom, respectively, for the present discussions).

\begin{figure}
\onefigure[width=7.5cm]{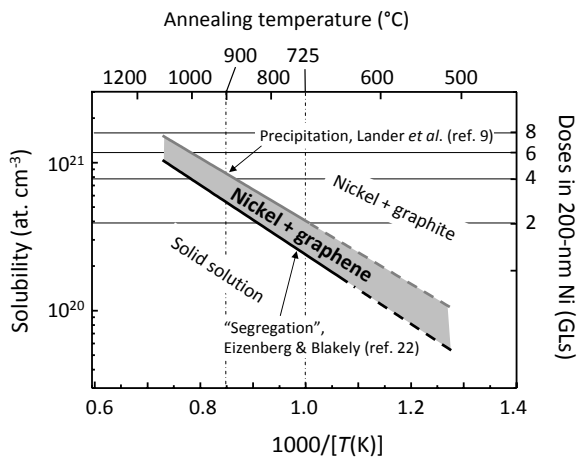}
\caption{Carbon solubility in nickel. In grey: range of concentration/temperature where graphene is at equilibrium on Ni surface. On the right-hand side, carbon concentration is expressed in terms of the corresponding dose one should implant in our 200-nm thick nickel films. One GL is the carbon atom density of one graphene layer (see text). The horizontal lines indicate the doses used in the present implantation experiments, and the vertical dotted lines correspond to the temperatures of the present annealing treatments.}
\label{fig.1}
\end{figure}

Blakely and co-workers deduced from their observations a heat of ''segregation'', $H_S$, which is 10 to 12 $\%$ \cite{Eizenberg1979} lower (higher in absolute value) than the above heat of precipitation $H_P$. These various considerations on ''segregation'' and precipitation are summarized by the Arrhenius plots of the two solubilities in fig.~\ref{fig.1}. In that figure, we have not directly used the heat of ''segregation'' derived by Eizenberg and Blakely\cite{Eizenberg1979}, as they have calibrated their carbon concentrations on the solubility data of Dunn \etal\cite{Dunn1968}, which are quite specific compared to several other works\cite{Lander1952, Hasebe1985}. We have thus re-calculated the ''segregation'' activation energy using the heat of precipitation $H_P$ given above \cite{Lander1952}. For the sake of simplicity and because of a large uncertainty on the derivation of the entropic pre-factor, we further assumed that this factor was equal to that for graphite precipitation. We have thus come up with a heat of graphene ''segregation'', $H_S$ $\sim$ $H_P$ - 11$\%$ $\sim$ -0.47 eV. Figure~\ref{fig.1} clearly shows the temperature range (gray region) in which, for a given carbon concentration, equilibrium thermodynamics will deliver monolayer graphene on conveniently oriented Ni surfaces. Actually, this inclination to produce an equilibrium graphene phase by segregation has been used in a recent study\cite{Liu2011}.

Now that we know the equilibrium states, let us recall that carbon atoms have a large diffusivity in Ni, which will influence the kinetics of precipitation. Carbon diffusivity $D_T$ at temperature $T$ reads: $D_T = D_0\exp(-E_D/kT)$ (in cm$^2$s$^{-1}$), where $D_0$ is an entropic pre-factor and $E_D$ the diffusion activation energy. Lander \etal, in their above mentioned study\cite{Lander1952}, have found $D_0$ =2.4818 cm$^2$s$^{-1}$, $E_D$ = 1.74 eV (their activation temperatures in kelvins have again been translated into activation energies in eV per atom). Carbon diffusion in Ni thus appears remarkably fast: at 725$^{\circ}$C (the lowest annealing temperature in our experiments), the diffusion length $L$ = $ 2\sqrt{D_T\tau}$ (where $\tau$ is the diffusion time) comes out at 1.2 $\mu$m for a 1-second anneal. Our nickel films being 0.2-$\mu$m thick, the carbon atoms will redistribute in the bulk of nickel whatever the treatment time. However, let us keep in mind that annealing will also induce recrystallization of the metal, which is a slower process\cite{Thiele2010}.

\section{Experimental: dose and anneal}
Let us now present the doses and treatments that we used for the present study. The most interesting aspect of ion implantation is that it allows one to set \textit{a priori} the maximum number of graphene layers that can be produced in a given process. Assuming that precipitation is uniform and only occurs on the surface of the metal film, this number simply equals the implanted dose divided by the graphene atomic layer density \textit{i.e.} $3.8\times10^{15}$ carbon atoms cm$^{-2}$. We call that elementary  dose a ''GL'' (graphene layer) in the following. The solubility of a given Ni film of thickness \textit{t} can be expressed in terms of dose with the simple formula $\delta = tS_T$. We used Ni films of thickness $t$= 200 nm ($2\times10^{-5}$ cm); now on, we write solubilities in GLs, after the expression $S$(GLs) = $S$(atoms cm$^{-3}$)$\times$ $2\times10^{-5}$/$3.8\times10^{15}$.

We have used a two-step annealing treatment: 900$^{\circ}$C for 30 min, followed by slow cooling ($\sim$0.5 Ks$^{-1}$) down to 725$^{\circ}$C and then quenching to $\sim$150$^{\circ}$C. Inferred from fig.~\ref{fig.1}, the carbon solubilities  $S_S$ for ''segregation'' and $S_P$ for precipitation, in our 200-nm thick nickel films, are respectively $S_S$ $\sim$2.5 GLs , $S_P$ $\sim$4 GLs at 900$^{\circ}$C and  $S_S$ $\sim$1 GL , $S_P$ $\sim$2 GLs at 725$^{\circ}$C.

The implanted carbon dose was varied between 2 and 8 GLs. Doses of 6 and 8 GLs were anticipated to deliver graphite precipitated at 900$^{\circ}$C, possibly under the form of few graphene layers, the 4-GL dose was anticipated to yield a mix of graphene and graphite precipitated during slow cooling to 725$^{\circ}$C; the 2-GL dose was anticipated to represent graphene ''segregation'' only. Additionally, we also applied a single 725$^{\circ}$C-anneal followed by quenching, to serve as a reference and check the impact on graphene formation of the important Ni recrystallization observed at 900$^{\circ}$C.

\section{Sample preparation details}
(1) 200 nm nickel thin films were e-beam evaporated on 300 nm thick silicon oxide thermally grown on silicon substrates. (2) Carbon ion implantation was performed at 80 keV in order to get a 100 nm projected range; the carbon atom distribution was simulated using the software developed by Ziegler\etal\cite{Ziegler2010}. Four implanted doses were used: 8$\times$10$^{15}$ (2 GLs), 1.6$\times$10$^{16}$ (4 GLs), 2.4$\times$10$^{16}$ (6 GLs) and 3.2$\times$10$^{16}$ atoms cm$^{-2}$ (8 GLs). (3) Annealing was performed by pushing the samples, hosted on a quartz boat into the furnace which had been preheated at the chosen temperature between 725 and 900$^{\circ}$C. (4) Quenching was obtained by the reverse process of taking the sample out of the furnace. 

Let us mention that our goal here being to analyse all growth mechanisms, and especially those involving crystal defects, we implanted carbon directly into the as-deposited Ni films, without any prior high temperature annealing, opposite to what is generally practised in the literature\cite{Reina2009, Garaj2010, Thiele2010, Chae2009}.

\section{Characteristics of films}
Graphene films were characterised using transmission electron microscopy (TEM): micrographs were recorded at 120 keV on a Topcon 002B microscope and at 300 keV using a Philips/FEI CM30. Plan-view TEM specimens were prepared by dissolving the nickel substrate and depositing the graphene on a TEM grid coated with a holey amorphous carbon film; cross-sections were prepared by tripod polishing and ion milling. Other characterisations, including Raman spectroscopy, electron backscatter diffraction (EBSD) of the Ni films, and electrical measurements, have been published in a previous paper\cite{Baraton2011}.

\begin{figure}
\onefigure[width=7cm]{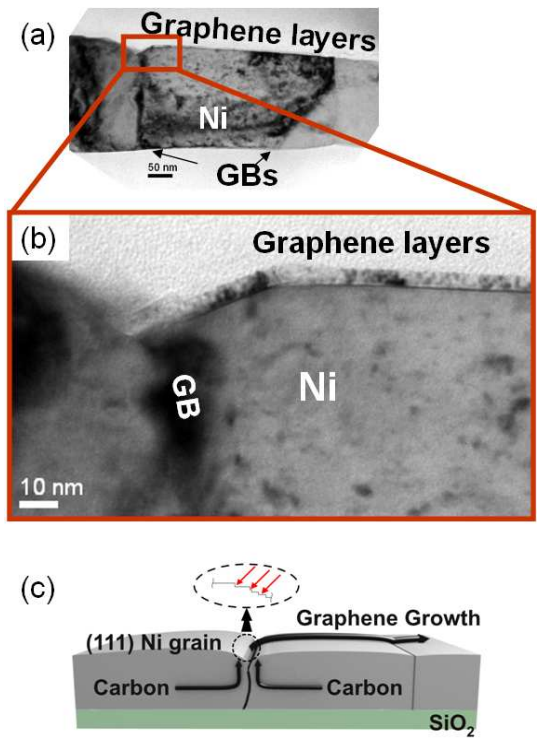}
\caption{TEM cross section of graphene or graphite on nickel grains. Sample implanted with the 4-GL dose, annealed at 900$^{\circ}$C for 30 min, slowly cooled to 725 $^{\circ}$C, then quenched. (a,b) TEM image showing the connection between a nickel grain boundary and graphene layers at the surface of the film. Note that graphene covers only one nickel grain, the left-hand grain remains bare. (c) Schematic representation of the probable nucleation and growth mechanism.}
\label{fig.2}
\end{figure}

Annealing appears to have modified the Ni films (fig.~\ref{fig.2}), inducing in particular substantial grain growth. Electron diffraction, consistent with EBSD (see ref.~\cite{Baraton2011}) indicates that the Ni grains are most often $<111>$ oriented. Large nickel areas are covered with a variable number of graphene layers. Figure~\ref{fig.2} shows a typical region where a few-layered graphene (FLG) film starts from the intersection with the surface of a nickel grain boundary (GB). Ni GBs obviously play a significant role in the mechanisms of nucleation and growth. However, the thickness of that surface graphite appears to vary over a wide range, always in the form of large good-quality crystals with the \textbf{c} axis perpendicular to the surface. The fact that we find areas with few graphene layers, others with thick graphite, or with no graphene, indicates that the in-plane carbon concentration, which was uniform after ion implantation, is strongly redistributed during annealing\cite{Baraton2011}.

\begin{figure}
\onefigure[width=8cm]{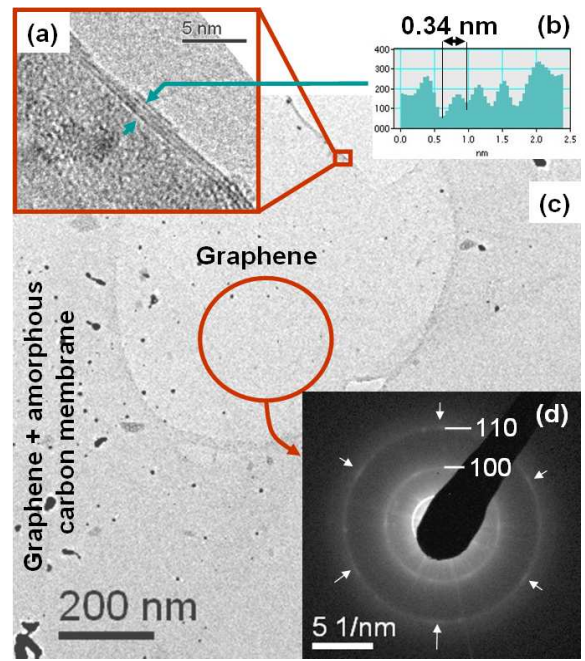}
\caption{TEM plan view of a graphene film after dissolving the supporting nickel layer and transferring the sample on a TEM grid. Sample implanted with the dose of 6 GLs, annealed at 900$^{\circ}$C for 30 min, slow cooling to 725 $^{\circ}$C, then quenched. (a) HRTEM image of the edge, where a local folding allows one to count the number of graphene layers; (b) Intensity profile of the image in (a), indicating a distance of 0.34 nm between the graphene layers. (c) Low magnification general view of the sample. (d) Selected area EDP (circle in (c)) exhibiting 100 and 110 graphene reflections with a distribution of orientations. A given orientation appears favoured as the diffracted intensity is enhanced with six-fold symmetry (arrows).}
\label{fig.3}
\end{figure}

\begin{figure}
\onefigure[width=7cm]{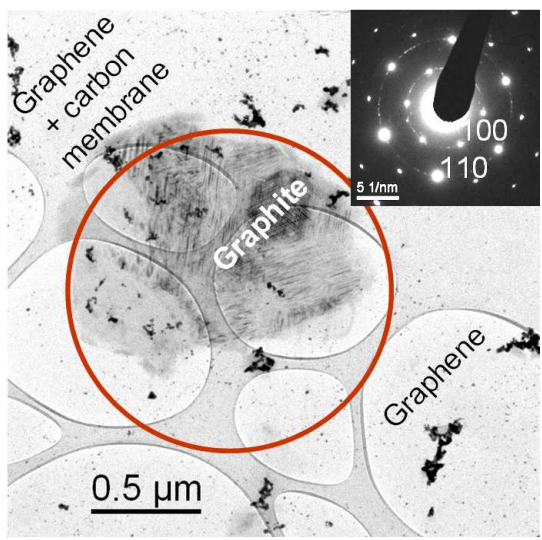}
\caption{TEM plan view of graphene and graphite flakes. Sample implanted with the 8-GL dose, annealed at 900$^{\circ}$C for 30 min, slowly cooled to 725 $^{\circ}$C, then quenched. The EDP of the circled area (inset) exhibits 100 and 110 graphite reflections with perfect single crystal order together with 100 and 110 rings; the former correspond to the large graphite crystal in the image and the latter to the surrounding nanocrystalline graphene.}
\label{fig.4}
\end{figure}

Plan-view observations are displayed in figs.~\ref{fig.3} and~\ref{fig.4}. The folding of the graphene sheet at the edge allows one to count the number of layers: three to four visible in figs.~\ref{fig.3}a,b. The structure does not significantly vary from one dose to the other: selected area electron diffraction (fig.~\ref{fig.3}d) indicates that the film is crystalline, but with very small disoriented grains, so that diffraction patterns are made of rings instead of spots. However, the intensity in those rings may be periodically reinforced with a six-fold symmetry (arrows in fig.~\ref{fig.3}d), which indicates the beginning of a long-range order in the associated areas (0.3 $\mu$m in diameter). The size of single-crystal domains can be directly evaluated from the inverse line width, through the Scherrer formula; it is of the order of 3.5 nm for grains participating in the long-range order (line width at the arrows in fig.~\ref{fig.3}d), and only 1.5 nm for measurements in between the arrows in fig.~\ref{fig.3}d. Therefore, the term “nanocrystalline graphene” should be more adapted to define the films investigated here.

On the other hand, thick and large graphite flakes are single crystals (fig.~\ref{fig.4}), which tends to indicate that their crystallization process is different (see below).

\section{Mechanisms of growth}
\subsection{Graphite crystals and few layered graphene}
The presence of graphite on the Ni surface is always coupled with that of grain boundaries (GBs) in the present observations, as if the graphite had nucleated along the lines defined by the intersections of GBs with the surface. This observation confirms the recent data of Zhang\etal ~who have compared precipitation of graphene on single- and poly-crystalline nickel\cite{Zhang2010}: the surface curvature associated with the presence of a GB causes nucleation of ''multilayer graphene''. The TEM observations (\textit{e.g.} fig.~\ref{fig.2}) indicate that graphite flakes have formed with the help of a large number of nickel atomic steps, in a geometry almost perpendicular to the side wall of a nickel grain. GBs are well-known preferential precipitation sites. However, the accumulation of carbon there suggests another role, which is that of drain for carbon atoms. That draining effect of GBs is most probably associated with their migration during annealing. The latter would provide a growth configuration which recalls that found with nanotubes, where the metal leaves the volume it occupied to the growing graphitic planes\cite{Lin2007, HeSoumis2011}. Thus, the development of graphite flakes is most probably correlated with GB migration. At this point, we note that when using a Ni film that has been recrystallized during a pre-thermal treatment, graphite growth appears indeed to be negligible\cite{Reina2009, Garaj2010, Chae2009}. We also note that when the surface topography is favorable (large radius of curvature of the side wall of the grain and small depression, thus creating few step layers), few layered graphene (FLG) films can be formed\cite{Baraton2011}.

Regarding the mechanism that brings the carbon atoms to sites where graphite/FLG is forming, let us first remark that the local carbon concentration, prior to annealing, is imposed by the particular implanted dose, which is always smaller than the number of atoms present in the many layers of graphene (fig.~\ref{fig.2}). Thus the carbon in that graphite has to come from distant sources. Let us then recall that in the case of carbon nanofibres, after pioneers in the field had published quite high values of the growth activation energy\cite{Baker1972}, more recent data\cite{Hofmann2005, Cojocaru} present lower values in the range 0.2-0.4 eV, indicating that in those cases, surface diffusion is the dominant way of transporting the carbon atoms that fuel the growth. Grain boundaries appear here to play the role of surfaces to fulfil the high carbon demand of the growing graphite/FLG flakes. Once a particular nucleation site is triggered and activated, it rapidly ''pumps out'' carbon atoms from the surrounding grains, thanks to grain boundaries. A diffusive flux of carbon atoms is thus established towards the nucleation site, leading to a strong redistribution of the carbon inside the nickel. As shown in fig.~\ref{fig.2}, once nucleated, graphene layers grow laterally from the particular nucleation site and cover the nickel surface, possibly crossing ''inactive'' GBs (\textit{i.e.}, GBs where there is no surface curvature - see fig.~\ref{fig.2}a, the GB on the right-hand side) as already observed recently\cite{Reina2009, Chae2009}. 

\begin{figure}
\onefigure[width=7.5cm]{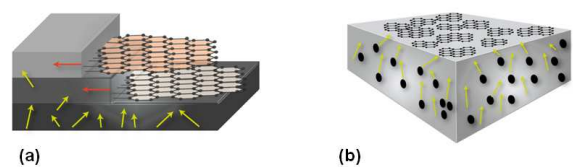}
\caption{Two types of growth processes starting from carbon atoms dissolved into Ni: (a) Lateral growth of high quality graphite/few layered graphene. (b) Growth of nanocrystalline graphene after surface precipitation of carbon atoms.}
\label{fig.5}
\end{figure}

Step flow gowth (fig.~\ref{fig.5}a) has indeed been observed for individual graphene layers, in the case of carbon nanofibres grown with Ni catalyst,\cite{Helveg2004, Abild-Pedersen2006, Hofmann2007} in the case of epitaxy of graphene on Ru(111)\cite{Marchini2007}, and in the nucleation of single-walled carbon nanotubes on Co\cite{Zhu2005}. The lateral growth that occurs here then recalls again the mechanism observed in multiwall carbon nanotube growth\cite{Rodríguez-Manzo2009, Lin2007, HeSoumis2011}, especially when the metal is in the form of nail-shaped particles, where the surface under the head of the nail delivers the graphene layers, and its body then guides them in the form of nanotube/nanofibres\cite{Lin2007, HeSoumis2011}. Such a scheme would explain why large graphite flakes (fig.~\ref{fig.3}) seem in some occurrences to have just replaced a certain thickness of nickel: expending nickel grains would offer room on their surface for graphite to grow laterally. 

So the less energetic path for graphite formation would be that summarized in fig.~\ref{fig.2}c, where grain boundaries accelerate the diffusion of carbon atoms toward nucleation sites, from which graphite/FLG films grow laterally (fig.~\ref{fig.5}a), as the GB drifts away. It is interesting to note that the graphite thickness in that case depends on the depth of the groove associated with the GB, and on the local surface curvature. As quoted above, FLG films are also synthesized according to the same mechanism (\textit{e.g.} fig.~\ref{fig.3} and ref. \cite{Baraton2011}), when the surface topography  is favorable, \textit{i.e.}, when the intersection of the grain boundary with the Ni surface exhibits a large radius of curvature and only a few atomic steps are available for graphene nucleation there. 

\subsection{Nanocrystalline graphene films}
The quasi absence of long-range order in our nanocrystalline graphene films (see the rings in the diffraction patterns; figs.~\ref{fig.3} and ~\ref{fig.4}) indicates that the mechanism at the origin of their existence is quite different from that discussed above: there is little correlation between the crystal orientation at one place and another. Thus, the nucleation rate in that case must have been extremely high, and it has obviously involved a very high density of nucleation sites, which is compatible with the fact that the Ni surface is far from perfect (see \textit{e.g.}, fig.~\ref{fig.2}). Thus also, this process has implied only local transport of carbon atoms (fig.~\ref{fig.5}b).  
An explanation is that the nanocrystalline graphene has most probably developed during quenching, below 725$^{\circ}$C, because of the very high supersaturation, associated with still significant carbon diffusivity\cite{Lander1952} (fig.~\ref{fig.5}b). Data from the literature indicate that such a disorder is indeed possible even in a constant-temperature regime, as for example in the case of C-segregation at the surface of platinum (100)\cite{Morgan1968}. The key word again is high supersaturation. Such conditions are essentially opposite to the unsaturated regime with which Blakely and co-workers observed single crystal graphene formation (see above)\cite{Eizenberg1979}. Our 2-GL dose, which was aimed at obtaining Blakely and co-workers$^{\prime}$ graphene ''segregation'',\cite{Shelton1974, Eizenberg1979, EizenbergJCP1979} thus did not deliver it. The precipitation that occurs during quenching is probably the main reason why we did not observe it; we always get more than one graphene layer. Another reason would be related to the very strong re-distribution of carbon that takes place throughout the Ni films during the present treatments. Whatever the exact cause of that redistribution, we did observe it with all the doses, including the 2-GL. Thus in reality, the local carbon concentration in the nickel was everywhere different from the concentration we imposed by implantation. As the carbon-concentration window for graphene ''segregation'' on Ni surface is quite narrow (fig.~\ref{fig.1}), the regions of sample where the concentration condition was fulfilled were probably very few.

Let us note, however, that nanocrystalline graphene is the form of graphene generally obtained when using low-cost growth techniques\cite{LeeCM2011, Turchanin2011, LeeCS2011}, and that its properties may be tuned so that it can be introduced in future devices such as chemical sensors.  

\section{Conclusions}
In conclusion, we have used ion implantation to study the growth mechanism of graphene layers on nickel thin films. Ion implantation appears as a versatile technique to study growth mechanisms, allowing controlled and predefined amounts of carbon to be introduced inside the nickel. Such control is of paramount importance for the understanding of the nucleation and growth of graphene on metal films. After having observed a strong in-plane carbon redistribution upon high-temperature annealing, we propose two modes of growth: (1) graphene nucleates at dedicated sites, where the local surface curvature of the Ni surface is strong, mostly along grain boundaries, and then grows laterally on the nickel surface, under the form of graphite crystals of varying thickness, some only a few layers thick; (2) graphene layers develop thanks to direct surface precipitation during quenching. Finally, we point out that, provided the thermodynamical conditions are carefully settled, and provided a proper metal surface can be used where the nucleation sites are controlled, ion implantation should be particularly adapted to graphene synthesis, since a dose of 3.8$\times$10$^{15}$ carbon atoms cm$^{-2}$ (easy to achieve) corresponds to the atomic density of a graphene layer.

\acknowledgments
We thank Dr. G. Rizza and Dr. P.-E. Coulon, LSI, Ecole Polytechnique, France, for the use of the CM30 TEM, and Dr. G. Garry and Dr. S. Enouz-Vedrenne (Thales R$\&$T France) for access to the Topcon 002B. This work has been supported by the Region Ile-de-France in the framework of C$^\prime$Nano IdF. C$^\prime$Nano IdF is the nanoscience competence center of Paris Region, supported by CNRS, CEA, MESR and Region Ile-de-France. Y.H. Lee and D. Pribat would like to acknowledge support from WCU program through the NRF of Korea, funded by MEST (R31-2008-000-10029-0).

\end{document}